# Polarization-selective branching of stop gaps in three-dimensional photonic crystals

Priya and Rajesh V. Nair*

Laboratory for Nano-scale Optics and Meta-materials (LaNOM)

Department of Physics, Indian Institute of Technology (IIT), Ropar

Rupnagar, Punjab 140 001 INDIA

*Email: rvnair@iitrpr.ac.in

**Abstract**

We study the direction- and wavelength-dependent polarization anisotropy in light scattering at the air-photonic crystal interface as a function of angle of incidence for TE and TM polarized light. This is done using optical reflectivity measurements at high-symmetry points in the Brillouin zone of a three-dimensional photonic crystal with *fcc* symmetry. Polarized reflectivity measurements indicate the presence of stop gap branching for TE polarization, which is absent for TM polarization till the Brewster angle at *K* point. In contrast, stop gap branching is present for both TE and TM polarizations at *W* point due to the intricate mixing of crystal planes. This characteristic behavior signifies the inevitable role of energy exchange in the stop gap branching. The measured polarization anisotropy shows a prominent shift in the Brewster angle for on-resonance wavelength as compared to the off-resonance along both *K* and *W* points, and that, in accordance with theory. Our results have implications in polarization-induced light scattering in sub-wavelength photonic structures like plasmonic crystals, and meta-materials.

# I. Introduction

The fundamental understanding of light-matter interactions in photonic meta-materials is very important to design them for useful applications in lasing, solid state lighting, and photon management in photovoltaic devices [1]. Photonic crystal structures belong to a class of meta-materials wherein the dielectric constant is spatially periodic in all three-orthogonal directions [2, 3]. Depending on the spatial period and the difference in dielectric constants, frequency gaps are formed in particular directions of light propagation known as photonic stop gaps [4]. The photonic stop gaps originate due to Bragg diffraction of light by photonic crystal planes. The stop gaps possess strong polarization dependent characteristics owing to the vectorial nature of light [2]. When the stop gaps in all directions occur at the same frequency range for different polarization states of incident light, a photonic band gap is formed [5]. The photon density of states is zero inside the photonic band gap which results in meticulous changes in spontaneous emission decay rates [6, 7]. Such control on the light emission process has potential applications in nano-lasers and quantum electrodynamics [8, 9]. It is a challenging task to fabricate photonic crystals evincing photonic band gap due to strict requirements of specific crystal symmetry and refractive index contrasts [10]. Therefore, much simpler photonic crystal structures, which possess only stop gaps, are largely explored. This has lead to the development of self-assembled three-dimensional (3D) photonic crystals with face centered cubic (*fcc*) symmetry that are made using colloidal suspension consisting of sub-micron spheres [4]. Self-assembled 3D photonic crystals are more attractive due its ease of fabrication, more versatility, and its robust functionalities [11]. However, they are excessively prone to implicit defects and disorder [4]. Fine tuning of synthesis conditions provide high quality photonic crystals with optical response analogous to theoretical predictions [12]. This provides a platform to investigate various exotic optical processes associated with 3D photonic crystals [13].

A magnificent optical process in two- or three-dimensional photonic crystals is the multiple Bragg diffraction that occurs when the tip of the incident wave vector spans a high-symmetry point in the Brillouin zone. This is assisted with branching of stop gaps in the optical reflectivity or transmission measurements [14-16]. Multiple Bragg diffraction in 3D photonic crystals has been a topic of intense research [17-22] and generally interpreted as the inherent property of photonic crystals using non-polarized light [19, 21-26]. Howbeit, formation of stop gap is a strong polarization-dependent process and hence polarization-resolved multiple Bragg diffraction is highly desirable. There are attempts made to map the polarized reflectivity spectra at different high-symmetry points in the Brillouin zone of photonic crystals with *fcc* symmetry [27-31]. However, a comprehensive understanding of polarization-induced stop gap branching in different symmetry directions and the role of energy exchange is yet to explore in 3D photonic crystals.

In this paper, we present rigorous experimental studies on the angle- and polarization-induced branching of stop gaps at *K* (*U*) and *W* points in the hexagonal facet of the Brillouin zone of crystals with *fcc* symmetry. We discuss, using the optical reflectivity spectra, the evolution of stop gaps at *K* and *W* points for TE and TM polarizations of light. The inflow of energy leading to the branching of stop gaps at certain incident angles is unequivocally explained using polarization-dependent measurements. The stop gap branching at *K* point results in two identical peaks for TE polarization whereas that at *W* point shows three peaks embedded in a complex reflectivity profile for both TE and TM polarizations. The polarization anisotropy factor is estimated to elaborate the direction- and wavelength-dependent Brewster effect in 3D photonic crystals. Our measured polarization anisotropy values are in complete agreement with the calculated values for the ideal crystals with *fcc* symmetry which signifies the superior quality of our photonic crystals.

## II. Samples and measurement geometry

Self-assembled 3D photonic crystals are grown on a glass substrate using convective self-assembly method [19, 22, 32]. We use commercially available polystyrene (PS) spheres (M/S. Microparticles GmbH) of diameter (*D*) 280 ± 6 *nm*. Scanning electron microscope (SEM) is used to visualize the structural ordering of photonic crystals. Angle- and polarization-dependent reflectivity measurements are done in the specular geometry using PerkinElmer Lambda 950 spectrophotometer. The sample is mounted in a way that enables us to access the *K* (*U*) and *W* points in the Brillouin zone of crystal [22, 31]. The source used is a Tungsten-Halogen lamp and a photomultiplier tube detector is employed to collect the reflected photons. The polarizer (Glan-Thomson) is mounted in the incident light path to select either TE or TM polarization of light. The polarization is defined with respect to the plane of incidence which is perpendicular to the top surface of the crystal. The beam dimensions on the sample are $5 \times 5$ *mm*. The average domain size (obtained from SEM images) of the sample is $100 \times 100$ *μm* and therefore more than 50 domains are probed from the crystal in the reflectivity measurements.

Figure 1(a) shows the SEM image of the photonic crystal that exhibits hexagonal packing of PS spheres on the surface. This represents the (111) plane of the crystal with *fcc* symmetry [33]. The image is captured around a vertical crack which is intrinsic to self-assembled photonic crystals [4]. The sample presents an excellent ordering on either side of the crack and also in the depth. Fig. 1(b) shows the 3D Brillouin zone of the crystal with relevant symmetry points and the lines joining them along which the propagation of light is examined in our work. The light is incident normally on the (111) plane to probe the photonic stop gap in the $\Gamma L$ direction and the crystal is illuminated at different angles of incidence ($\theta$) to access other high-symmetry points ($K$, $U$, and $W$) [31]. The diffracted wavelength ($\lambda_{hkl}$) from planes with Miller indices ($hkl$) in photonic crystals can be calculated using the Bragg's law for optical diffraction given as [30];

$$\lambda_{hkl} = 2 \times n_{eff} \times d_{hkl} \times \cos\left(\alpha - \sin^{-1}\left(\frac{1}{n_{eff}}\sin\theta\right)\right) \quad\quad\quad\quad\quad\quad (1)$$

where $d_{hkl}$ is interplanar spacing, $n_{eff}$ is the effective refractive index, and $\alpha$ is internal angle between the ($hkl$) and (111) plane.

Figure 2 shows the calculated stop gap wavelengths along the $LK$ and $LU$ lines in the cross-section of the Brillouin zone (inset). The (111) [solid line] stop gap shifts towards the shorter wavelength region whereas the ($\bar{1}11$) [dashed line] and (200) [dotted line] stop gaps show an opposite dispersion with increase in $\theta$. The (111) stop gap crosses the ($\bar{1}11$) and (200) stop gaps at the same angle ($\theta = 56°$) due to the equal length of $LK$ or $LU$ on the hexagonal facet of the Brillouin zone [31]. The formation of stop gaps at $K$ ($U$) points can be well understood through evoking the Laue diffraction condition as shown in Fig. 2 (inset) [14]. At a certain value of $\theta$, the incident wave vector passes through the $K$ point, then the Laue condition is satisfied simultaneously for reciprocal lattice vectors corresponding to (111) and ($\bar{1}11$) plane with conditions: $\vec{k}_0 + \vec{G}_{111} = \vec{k}_1$ and $\vec{k}_0 + \vec{G}_{\bar{1}11} = \vec{k}_2$. Here, $\vec{k}_1$ and $\vec{k}_2$ are diffracted wave vectors corresponding to the reciprocal lattice vectors $\vec{G}_{111}$ and $\vec{G}_{\bar{1}11}$, respectively. The interaction of three reciprocal lattice vectors ($\vec{G}_{000}$, $\vec{G}_{111}$, and $\vec{G}_{\bar{1}11}$) engenders multiple Bragg diffraction at the $K$ point. A similar set of diffraction conditions can also be applied at the $U$ point.

There persists a subtle issue on assigning the crystal planes responsible for the origin of stop gaps when the wave vector spans the *K* or *U* point. High resolution microscope images can be used as a gauge for identifying the orientation of crystal planes involved in the formation of stop gaps at *K* or *U* point [19, 30]. This is done by imaging the samples across the depth which reveals either hexagonal or square ordering of spheres comprising the {111} or {200} family of planes, respectively, in crystals with *fcc* symmetry. However, it can give specious results as the assignment of crystal planes must come from optical spectroscopic methods. Therefore, the optical reflectivity measurements are more consistent since large numbers of domains with many crystal planes deep into the sample are taken into account.

## III.     Optical reflectivity

### A.  Along $\Gamma L$ direction

The reflectivity (transmission) spectra at near-normal incidence ($\theta = 10°$) show a peak centered at 610 *nm* with a reflectivity (transmittance) of 55% (2%) which constitutes the signature of (111) stop gap in the $\Gamma L$ direction. The measured photonic strength is 5.75% which is in agreement with calculated values from the photonic band structure [34]. The Bragg Length ($L_B$) that dictates the length of light attenuation at the stop gap wavelength is estimated to be 2.4 *μm* or $10d_{111}$, where $d_{111}$ is the interplanar spacing in the [111] direction. The thickness (*t*) [35] obtained from the Fabry-Perot (F-P) fringes, in the long wavelength limit, is 9 *μm* (~35 ordered layers) or $t = 3.9L_B$. This indicates that the crystals are strongly interacting and the finite-size effects are minimized in the direction of propagation.

### B.  Along $\Gamma K$ (*U*) direction

Figure 3(a) depicts the reflectivity spectra measured with TE polarized light when tip of the wave vector shifts towards the *K* (*U*) point in the hexagonal facet of Brillouin zone of crystal with *fcc* symmetry. The reflectivity spectra are shown for selected values of $\theta$ on either side of the expected stop gap crossing (see Fig. 2). The measured spectra show the (111) stop gap at 539 *nm* with a reflectivity of 52% for $\theta = 45°$ (black dotted). A new peak also arises at 490 *nm* with low reflectivity near the short-wavelength band edge. The (111) stop gap is blue shifted with slight reduction in reflectivity whereas the new peak is red shifted with enhanced reflectivity for $\theta < 56°$. At $\theta = 56°$ (red solid), the spectra shows a remarkable feature wherein both peaks show equal reflectivity (~40%) and line-width (~20 *nm*). Here at $\theta = 56°$, both peaks are trying to diffract at the same wavelength that results in an avoided crossing with exchange in their spectral positions. The (111) stop gap is centered at 495 *nm* and the new peak is at 526 *nm* with 31 *nm* separation which is higher than the individual line-width in correlation with strong-coupling regime [16]. For $\theta > 56°$, the (111) stop gap acquires its intensity with further blue shift whereas the new peak diminishes beyond the crossing retaining the red shift. It is fascinating to observe the simultaneous diffraction in the form of multiple peaks for $45° \leq \theta \leq 65°$. Such multiple Bragg diffraction is associated with the branching of stop gaps in the reflectivity spectra and extends over an angular range of more than 20°. This is the largest angular range of stop gap branching reported in literature to the best of our knowledge.

Figure 3(b) depicts the reflectivity spectra at different $\theta$ for TM polarization. Contrary to TE polarization, the peak reflectivity of (111) stop gap constantly decreases from 28 to 4% accompanied with narrowing of the line-width for change in $\theta$ from 45° to 59°. The observed closing of gap is similar to the theoretical calculation for TM photonic bands [31]. However, the photonic strength is nearly the same (~4%) for $45° \leq \theta \leq 59°$. The decrease in peak reflectivity with constant photonic strength at higher $\theta$ promulgates that the deterioration of stop gap is a polarization-induced process and not due to any structural imperfections in crystals. Fig. 3(c) shows (111) stop gap at 491 *nm* with a minimum reflectivity of 2.5% for $\theta = 62°$ (solid line). The monotonous decrease in the reflectivity values for $\theta \leq 62°$ is attributed to the Brewster angle ($\theta_B$) effect at the air-crystal boundary. For $\theta > 62°$, the (111) stop gap reflectivity increases gradually; albeit small. A small reflectivity peak is visible at 537 *nm* which is having a red shift for $\theta \geq 62°$ (shown using an arrow in Fig 3(c)). Unlike the case of TE polarized light, the (111) stop gaps for TM polarized light do not show anti-crossing and branching of stop gaps. Thus, there exists a substantial difference in the interaction of TM polarized light with photonic crystal structure as compared to TE polarized light.

### C. Along the *ΓW* direction

Fig. 4(a) shows the TE polarized reflectivity spectra at selected $\theta$ for wave vectors shifting towards the *W* point. At $\theta = 50°$, the (111) stop gap is at 526 *nm* with peak reflectivity of 53%. At $\theta = 54°$, in addition to the (111) stop gap at 518 *nm*, we observe the origin of two weak reflectivity lobes at 462 *nm* ($S_1$ peak) and 482 *nm* ($S_2$ peak). The (111) stop gap with reduced intensity is blue shifted whereas the $S_1$ and $S_2$ peak become well-resolved showing a red shift for $58° \leq \theta \leq 64°$. It is commendable to observe the anti-crossing of (111) stop gap with $S_2$ peak at $\theta = 65°$ (red solid). The (111) stop gap is now centered at 483 *nm* and the $S_2$ peak is centered at 506 *nm* with nearly equal peak reflectivity. The (111) stop gap and $S_2$ peak tend to diffract at the same wavelength leading to an avoided crossing behavior. The $S_1$ peak does not have any noticeable shift with increase in $\theta$. The (111) stop gap maintains the blue shift and the anti-crossed $S_2$ peak remains intact at 506 nm for $\theta = 65°$ to 67°.

Figure 4(b) and 4(c) show the reflectivity spectra at different $\theta$ for TM polarized light. The (111) stop gap is shifted from 525 to 505 *nm* with notable decrease in peak reflectivity for change in $\theta$ from 50° to 58°; see Fig. 4(b). Additionally, at $\theta \geq 56°$, two very weak reflectivity lobes are also observed at 477 *nm* ($S_3$ peak) and 493 *nm* ($S_4$ peak). At $\theta = 59°$ (red solid), an appealing feature is observed wherein the (111) stop gap anti-crosses the $S_4$ peak. The (111) stop gap is now centered at 493 *nm* with minimum reflectivity (~6%) due to Brewster effect and the $S_4$ peak is at 503 *nm*. The (111) stop gap continues a blue shift to 490 *nm* with enhanced reflectivity of 8% for change in $\theta$ from 61° to 63° as seen in Fig. 4(c). The $S_3$ peak with increased reflectivity still remains at 477 *nm* for $\theta = 61°$. In analogy to TE polarization at $\theta = 65°$, the (111) stop gap anti-crosses the $S_3$ peak with both having an equal reflectivity (~8%). The (111) stop gap continues its blue shift with higher reflectivity and the $S_3$ peak is red shifted to 491 *nm* with lower reflectivity for $\theta = 67°$.

## IV. Analysis of optical reflectivity

### A. Estimation of $n_{eff}$ and $D$

The calculation of $\lambda_{hkl}$ using eq. (1) requires the value of $n_{eff}$ and $D$ and their estimation is quite delicate in photonic crystals. Many models are used to estimate the value of $n_{eff}$ such as those using material refractive indices with aided knowledge of their filling fractions or techniques based on spectroscopic ellipsometry [36]. However, the $n_{eff}$ can be estimated in a unique way using the measured optical reflectivity spectra [37]. Let us assume that the stop gap branching occurs at $\theta_K = 56°$ for wave vector incident along the *K* point [see Fig. 2 (inset)]. Using this geometry and Snell's law, we can estimate $n_{eff} = \sqrt{3}\ sin\theta_K = 1.436$. Also, at $\theta_K = 56°$, the reflectivity spectra [Fig. 3(a)] show a trough at $\lambda_K = 510$ *nm* that results in the high transmission of light. The photonic crystal structure acts like a homogenous medium at $\lambda_K$ with certain $n_{eff}$ and the light propagation is well described using the free-photon dispersion relation. Therefore, the relation between the frequency ($\omega$) and the wave vector ($\vec{k}$) is written as;

$$\omega = \frac{c\vec{k}}{n_{eff}} = \frac{c|\overrightarrow{\Gamma K}|}{n_{eff}} \ldots\ldots\ldots\ldots\ldots\ldots\ldots\ldots (2)$$

where $|\vec{\Gamma K}|$ represents the length of incident wave vector at $\theta_K$ and $c$ is the speed of light. Rewriting eq. (2) in-terms of $\lambda_K$ using $|\vec{\Gamma K}|$ and $\omega$ $(= 2\pi c \lambda_K^{-1})$, we can estimate the value of $D$ as:

$$D = {3\lambda_K}/{4n_{eff}} \ldots\ldots\ldots\ldots\ldots\ldots\ldots\ldots (3)$$

The obtained value of $D$ is 266 $nm$ for the light incident along $K$ point. If we assume that the stop gap branching occurs at $U$ point, then we estimate $n_{eff}$ = 1.805 and $D$ = 195 $nm$. Such large variations in the value of $n_{eff}$ and $D$ cannot be justified. Using the values of $n_{eff}$ and $D$ along $K$ point, the stop gap wavelength in the $\Gamma L$ direction is calculated to be 623 $nm$; in close agreement with the measured value (= 610 $nm$) at $\theta$ = 10°. We have also calculated the stop gap wavelength using eq. (2) along the $\Gamma W$ direction as 483 $nm$ which is consistent with the measured value (= 493 $nm$) at $\theta$ = 65° for $W$ point. This assures the credibility of using above parameters in interpreting the stop gap branching at $K$ and $W$ points as discussed below.

## B. Along $\Gamma K$ direction

Figure 5(a) and 5(b) show the calculated diffraction wavelengths using eq. (1) for planes, such as, (111) [solid line], ($\bar{1}$11) [dashed line], and (200) [dotted line] and measured stop gap wavelengths (symbols) for TE and TM polarizations, respectively. The value of $\alpha$ used is 70.5° for the ($\bar{1}$11) plane and 54.7° for the (200) plane. The measured TE and TM stop gap wavelengths are identical for $\theta \leq 45°$ ratifying that the value of $n_{eff}$ is same for both polarizations. The measured (squares) and calculated (111) stop gap wavelengths are in good agreement for both TE and TM polarizations. When the measured (111) stop gap appears near the crossing regime, it deviates from its calculated curve due to the band repulsion forced by the presence of new peak as seen in Fig. 5(a). The new peaks (circles) emanate near crossing regime, are in good agreement with calculated ($\bar{1}$11) diffraction wavelengths. Our measured band crossing occurs at 0.75 $a/\lambda$ ($a$ is the *fcc* lattice constant) in complete agreement with calculated photonic band structure in the *Γ-L-K* orientation [31]. The measured (111) and ($\bar{1}$11) stop gaps show an opposite dispersion beyond crossing angle (=56°) in parallel with calculations. Fig. 5(b) clearly indicates the absence of any new peak except the (111) stop gap in the crossing regime for TM polarization. However, we observe a new peak far-off the calculated band crossing for $\theta \geq 62°$ which is in good agreement with ($\bar{1}$11) stop gap. The origin of ($\bar{1}$11) stop gap beyond $\theta_B$ (=62°) is in complete agreement with calculated reflectivity spectra for ideal photonic crystals with *fcc* symmetry [17].

### C. Along *ΓW* direction

Figure 6(a) and 6(b) show the calculated (line) and measured stop gap wavelengths (symbols) at different $\theta$ for TE and TM polarized light, respectively. The calculations are performed for (111) plane as this is the only plane which can be analyzed with certainty at *W* point. The ($\bar{1}$11) and (200) planes intercept the (111) plane at *W* point in a complex manner and therefore the value of $\alpha$ in eq. (1) cannot be deduced correctly leading to erroneous results in the calculations. The measured (squares) and calculated (111) stop gap wavelengths are in good agreement till $\theta \leq 58°$ for both TE and TM polarizations. The measured (111) stop gap is deviated from the calculated curve for $59° \leq \theta \leq 64°$ due to the appearance of $S_2$ peak (circles) as seen in Fig. 6(a). The (111) stop gap and the $S_2$ peak exhibit an avoided crossing at $\theta = 65°$ and thereafter have an opposite dispersion. The $S_1$ peak (triangles) originated at $\theta = 50°$ shows a slight red shift with increase in $\theta$. The $S_1$ and $S_2$ peaks arise due to the diffraction from {111} and {200} family of planes at the *W* point. Fig. 6(b) depicts the crossing of (111) stop gap with $S_4$ peak (diamonds) and $S_3$ peak (circles) at $\theta = 59°$ and $\theta = 65°$, respectively and beyond crossing both show an opposite dispersion. The *W* point is accessed at higher values of $\theta$ since the length *LW* is larger than the length *LK* on the hexagonal facet of Brillouin zone.

### D. Polarization anisotropy

The polarization anisotropy factor is estimated to comprehend the interaction of polarized light with photonic crystals. The anisotropy factor ($P_a$) is defined as: $P_a = R_{TM}/R_{TE}$; where $R_{TE}$ ($R_{TM}$) is the TE (TM) polarized reflectivity value. Fig. 7(a) and 7(b) illustrate $P_a$ at different $\theta$ for an off-resonance (squares) wavelength at 700 *nm*, and the on-resonance (circles) wavelength corresponding to (111) stop gap along the *LK* and *LW* line, respectively. The measured $P_a$ values are compared with the calculated values (triangles) obtained using Fresnel equations [38] for a film of refractive index of 1.436. The measured on- and off-resonance and calculated $P_a$ values are same at $\theta = 10°$ as there is no distinction between TE and TM polarizations at near-normal incidence. The $P_a$ value decreases with increase in $\theta$ for both on- and off-resonance wavelengths. The off-resonance $P_a$ value achieves its minimum at $\theta = 55°$ similar to the calculations. The minimum $P_a$ value corresponds to $\theta_B$ for an air-dielectric boundary with $n_{eff}^B$ as 1.428. This $n_{eff}^B$ is very close to the value of $n_{eff}$ obtained from the reflectivity measurements in Sec. IVA. The minimum $P_a$ value for on-resonance is achieved at $\theta = 62°$ which is higher than the off-resonance condition for the *LK* line as seen in Fig. 7(a). The variation in $\theta_B$ for on-resonance $P_a$ values is also confirmed by the presence of ($\bar{1}11$) stop gap (diamonds) beyond $\theta_B$ as seen in Fig. 7(a) (inset). Fig. 7(b) indicates the minimum on-resonance $P_a$ value at $\theta = 59°$, along the *LW* line. This shift in on-resonance $\theta_B$ is similar to the case of wave vectors shifting towards the *K* point. Fig. 7(b) (inset) shows the TM polarized reflectivity spectra for *K* point (solid line) and *W* point (dotted line) at $\theta = 62°$. The (111) stop gap for both symmetry directions is centered at 490 *nm* with lower reflectivity values for *K* point as compared to *W* point. This verifies the intrusive direction-dependent light reflection in accordance with calculations done for ideal photonic crystals with *fcc* symmetry [17].

## V. Discussion

The physical origin of stop gap branching can be quantitatively explained using the energy exchange taking place within the crystal. As seen in Sec. III, the (111) stop gap reflectivity value decreases and that for ($\bar{1}$11) stop gap increases with increase in $\theta$ till they become equal at $\theta$ = 56° for TE polarization. For $\theta > 56°$ the peak reflectivity of (111) stop gap increases slightly whereas that for the ($\bar{1}$11) stop gap decreases. This specifies a continuous exchange of energy between the (111) and ($\bar{1}$11) planes. Conversely, for TM polarization, the (111) stop gap reflectivity incessantly decreases to a minimum value, for $\theta \leq \theta_B$ (=62°) due to the Brewster effect. Hence, there is no sufficient light diffraction by (111) planes within the crystal to supply energy to ($\bar{1}$11) planes for $45° \leq \theta \leq 62°$. The reflectivity of (111) stop gap slightly increases for $\theta \geq \theta_B$ transferring energy to ($\bar{1}$11) stop gap. This validates the requirement of energy exchange for the stop gap branching. The decrease or increase in the peak reflectivity values with $\theta$ is also observed at *W* point for TE polarization as a consequence of energy exchange. However, at *W* point the TM polarized light efficiently excites multiple reflectivity peaks as it penetrates deep into the crystal due to complex interaction of {111} and {200} family of planes at *W* point. It is also noteworthy that the minimum value of stop gap reflectivity at $\theta_B$ is higher for the *W* point as compared to that at *K* point which suggests the direction-dependent polarization effects in photonic crystals.

Another interesting feature observed in our work is the band repulsion that supervenes near the crossing regime. It can be seen in Fig. 3(b) (inset) that at crossing angle (=56°), the TM polarized (111) stop gap appears at 510 *nm* which is exactly midway between (111) and ($\bar{1}$11) stop gaps for TE polarized light. This supports our earlier hypothesis that the (111) stop gap is repelled by the presence of ($\bar{1}$11) stop gap for TE polarized light whereas for TM polarized light the (111) stop gap smoothly follows the calculated pattern as the ($\bar{1}$11) stop gap is absent to enforce the band repulsion. Also at $W$ point, the anti-crossing of (111) stop gap is observed at $\theta = 65°$ for TE polarized light whereas that occur twice for TM polarization at $\theta = 59°$ and $65°$, respectively. The band repulsion is not obvious in the case of TM polarization may be due to the mixing of planes as evident from the reflectivity spectra with multiple crossings.

In a recent theoretical study, authors performed detailed simulations of reflectivity spectra of 3D photonic crystals with *fcc* symmetry using the effective medium approximation [39]. Their results reveal a single stop gap for TE polarization at all values of $\theta$ and vanishing of stop gaps at certain $\theta$ for TM polarization in contrast to our experimental results. Thus, our results suggest that interpreting the optical response of 3D photonic crystals using one-dimensional effective medium theory is highly questionable. The results shown in our work contemplate the direction-dependent optical response of 3D photonic crystals which cannot be explained using one-dimensional effective medium approximation.

The most enticing result of our work is the direction- and wavelength-dependent shift of $\theta_B$ in photonic crystals as shown in Sec. IVD. The minimum $P_a$ value is far above zero for on-resonance as compared to off-resonance or calculated $P_a$ values due to the unique light scattering from photonic crystals as shown in earlier theoretical calculations [40]. The observed shift in $\theta_B$ at on-resonance wavelength is in accordance with the calculated deviation of $7°$ for the ideal *fcc* crystals along the *LK* line [17]. Such closeness in the deviation of $\theta_B$ is demonstrated experimentally here for the first time. This polarization anisotropy is due to the involuted light scattering at the air-crystal boundary rather than a simple air-dielectric interface. The shift in $\theta_B$ is observed as 4° along the *LW* line and also the minimum $P_a$ value at on-resonance is higher at the *W* point as compared to the *K* point. This supports that the polarization anisotropy is dependent on high-symmetry points in photonic crystals similar to theoretical calculations [17]. This also upholds the fact that conventional definition of $\theta_B$ should be revisited in photonic crystals. The polarization anisotropy is peculiar to sub-wavelength meta-surfaces which is also shown in a recent work on monolayer of silica spheres [41].

The present work has significant impact on the prospective applications using photonic crystals such as the control of spontaneous emission leading to mirror-less lasing and solid state lighting [8, 9]. Most of these applications are performed with non-polarized excitation sources and the results are analyzed without imposing much attention to the vectorial nature of light. The intrusive interaction of polarized light as discussed in present work suggests that the applications like lasing in photonic crystals must be polarization-dependent. It is an open avenue to study the spectral and temporal dynamics of light emission at the crossing regime and how the emission is modulated at high-symmetry points in photonic crystals. The results also have prospects in the optics of plasmonic structures, two-dimensional array of spheres, and in meta-materials as the polarization of incident light play a vital role in exciting resonant modes in these structures.

## VI. Conclusions

In summary, we have shown a detailed study of direction- and polarization-dependent stop gap branching at high-symmetric points in photonic crystals with *fcc* symmetry using optical reflectivity measurements. The branching of photonic stop gaps when the tip of the wave vector spans the *K* point for TE and TM polarized light is analyzed. The reflectivity values of (111) and ($\bar{1}11$) stop gaps are exchanged with $\theta$ except at 56° wherein they have equal reflectivity values accompanied with avoided crossing of stop gaps for TE polarization. In contrast, TM polarized reflectivity spectra do not show stop gap branching and rather the reflectivity values of (111) stop gap keeps on reducing for $\theta \leq \theta_B$. The stop gap branching occurs at $\theta > \theta_B$ which results in the formation of ($\bar{1}11$) stop gap assisted with increase in the (111) stop gap reflectivity. The reflectivity spectra measured for wave vectors shifting towards the *W* point show branching of stop gaps into three peaks for both TE and TM polarized light. The anti-crossing of stop gaps is observed at 65° for TE polarization whereas that for TM polarization is observed at 59° and 65° which convinces the complex interaction of light at *W* point. The polarization-dependent stop gap mapping confirms the role played by the energy exchange in the stop gap branching at high symmetry points which is in fact also direction-dependent in photonic crystals.

We have also estimated the polarization anisotropy factor which attributes to the modification of $\theta_B$ in photonic crystals owing to the critical definition of $n_{eff}$. The observed shift in $\theta_B$ of 7° and 4° at on-resonance wavelength as compared off-resonance wavelength along the *K* and *W* points, respectively which is in complement to theoretical calculations. This deviation in $\theta_B$ is mainly due to the air-photonic crystal interface rather than air-dielectric interface as it is done in conventional optics. Our results establish the strong polarization anisotropy which is direction- and wavelength-dependent in photonic crystals and therefore it has vivid implications in designing photonic crystal based applications like low-threshold nano-lasers, solid state lighting, and wavelength filters.

**Acknowledgements**

The authors would like to thank the use of UV-Vis-NIR spectrophotometer, central facility at IIT Ropar and DST-SERB, Govt. of India for project [SB/FTS/80/2014] support. Priya thanks IIT Ropar for the PhD fellowship.

# Figures

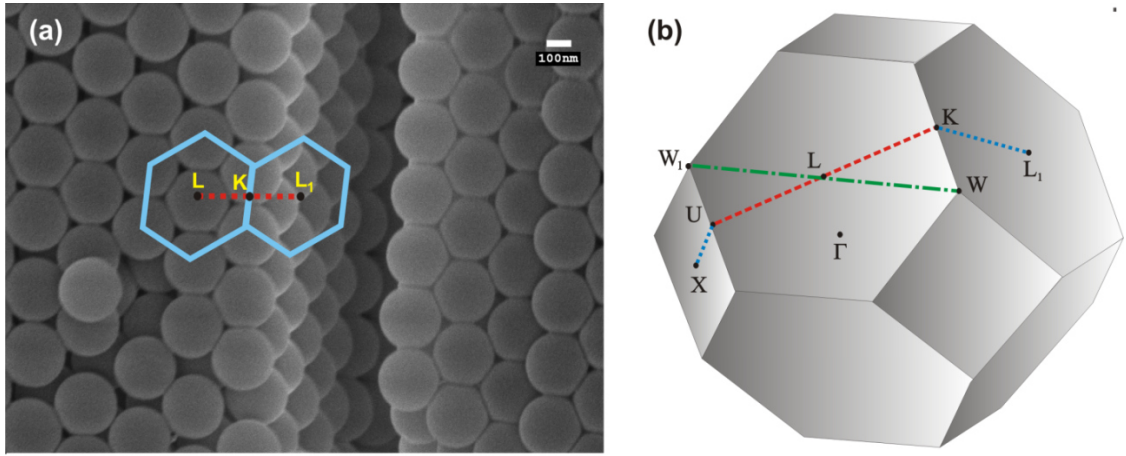

**Fig 1.** (a) The microscope image shows the hexagonal ordering of spheres on the surface that represents the (111) plane of the crystal with *fcc* symmetry. The image is captured across the crack to show fine structural quality in the depth of the sample. The hexagonal ordering across the depth indicate the ($\bar{1}11$) plane. (b) The 3D Brillouin zone of the crystal with the relevant symmetry points and the lines joining them.

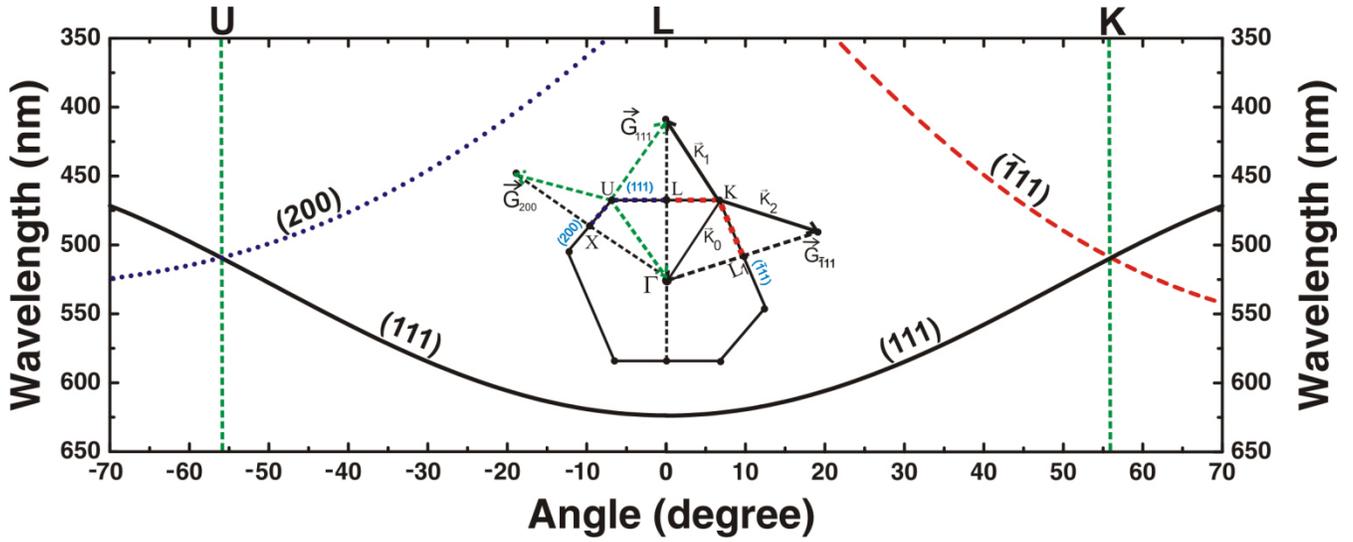

**Fig 2.** The calculated diffraction wavelengths showing stop gaps dispersion for wave vector shifting along the *LK* and *LU* lines on the hexagonal facet of Brillouin zone of the crystal with *fcc* symmetry. The calculations are done for different crystal planes relevant to our experiment using estimated values of $n_{eff}$ =1.436 and $D = 266$ *nm*. **Inset:** The cross-section of the Brillouin zone of crystal with *fcc* symmetry. The incident wave vector is $\vec{k}_0$ and $\vec{k}_1$, $\vec{k}_2$ are the diffracted wave vectors from the (111) and ($\bar{1}$11) planes, respectively. The wave vector $\vec{k}_0$ is shifted along the *K* point (red dashed line) which satisfies the Laue condition simultaneously for the reciprocal lattice vectors $\vec{G}_{111}$ and $\vec{G}_{\bar{1}11}$ leading to multiple Bragg diffraction at *K* point. Similarly, multiple Bragg diffraction can also occur at *U* point for the reciprocal lattice vectors $\vec{G}_{111}$ and $\vec{G}_{200}$.

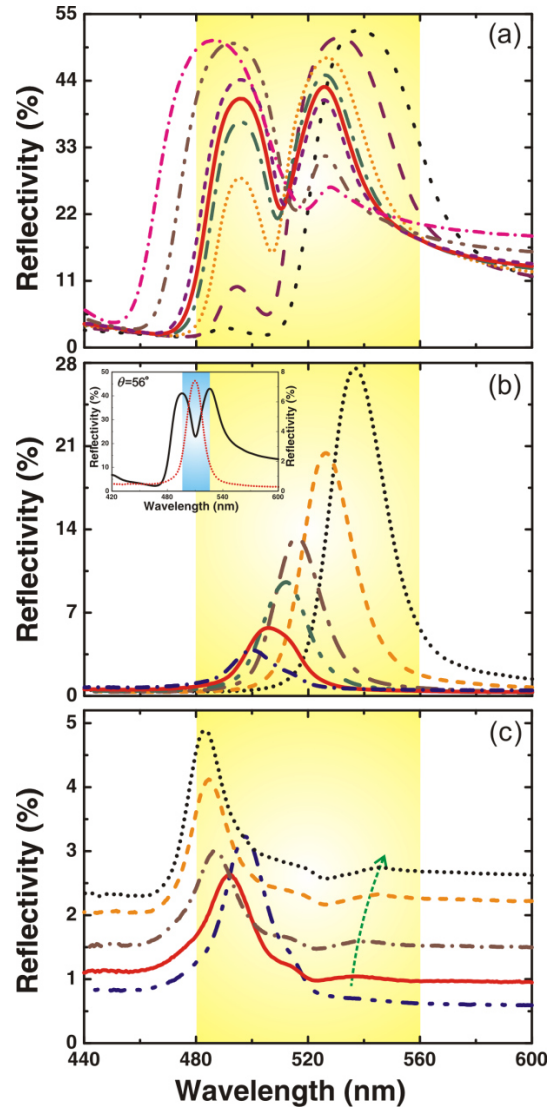

**Fig 3.** Reflectivity spectra measured for wave vectors shifting towards *K* point. (a) The measurements are done using TE polarized light at $\theta$ = 45° (dotted), 49° (dashed), 53° (short-dotted), 55° (dash-dot), 56° (solid), 57° (short-dashed), 61° (dash dot-dot), and 67° (short dash-dotted). The spectra show branching of stop gaps with equal intensity at $\theta$ = 56°. (b) The measurements are done for TM polarized light at $\theta$ = 45° (dotted), 49° (dashed), 53° (dash-dot), 55° (dash dot-dot), 57° (solid) and 59° (short dash-dot). The peak reflectivity value reduces rapidly with increase in $\theta$ and no stop gap branching is observed at any $\theta$ value. The inset shows the reflectivity spectra at $\theta$ = 56° for TE (solid) and TM (dotted) polarized light. (c) The measurements at $\theta$ = 60° (dash dot-dot), 62° (solid), 64° (short dash-dot), 66° (short dash), and 67° (dotted) also using TM polarized light. The intensity of the (111) peak increases from $\theta$ = 62° onwards, in addition to the building up of a weak reflectivity lobe in the long wavelength side (shown with an arrow).

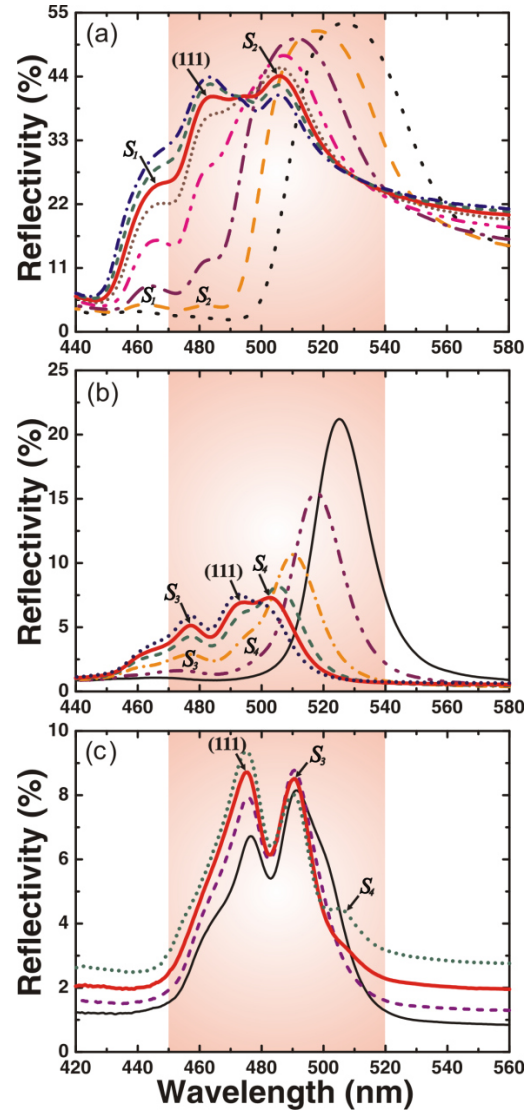

**Fig 4.** The measured reflectivity spectra for wave vectors shifting towards the *W* point for (a) TE polarization, and (b, c) for TM polarization. **(a)** The measurements are shown for $\theta = 50°$ (dotted line), 54° (dashed line), 58° (dash-dot line), 62° (dash dot-dot line), 64° (short dotted line), 65° (solid line), 66° (short dashed line), and 67° (short dash-dot line). The (111) stop gap anti-crosses the $S_2$ peak at 65° and the $S_1$ peak remains at the short-wavelength band edge without any anti-crossing. **(b)** The TM polarized reflectivity spectra for $\theta = 50°$ (thin solid line), 53° (dash-dot-dot line), 56° (dash-dotted line), 58° (dashed line), 59° (thick solid line), and 60° (dotted line). The (111) peak reflectivity decreases with increase in $\theta$. The (111) stop gap anti-crosses the $S_4$ peak at 59°. **(c)** The measurement at $\theta = 61°$ (thin solid line), 63° (dashed line), 65° (thick solid line), and 67° (dotted line). The (111) stop gap anti-crosses the $S_3$ peak at $\theta = 65°$.

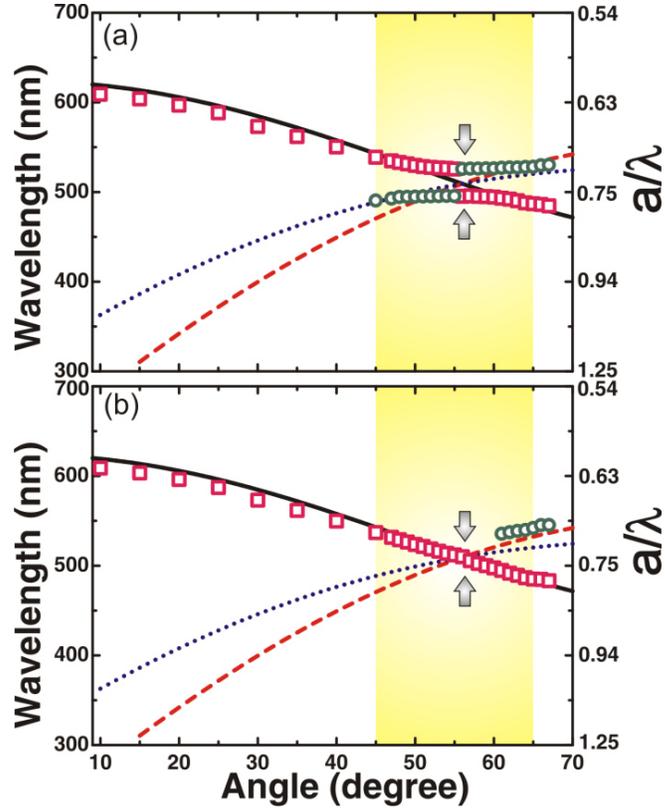

**Fig 5.** The measured (symbols) and calculated (lines) Bragg wavelengths for crystal planes (111) (solid), ($\bar{1}$11) (dashed), and (200) (dotted) that are intersecting at the *K* point for (a) TE and (b) TM polarized light. The measured (111) stop gap (squares) wavelengths are in good agreement with calculated values. The second reflectivity peak (circles) is in agreement with calculated ($\bar{1}$11) stop gap wavelengths. (a) When the (111) stop gap approaches the crossing regime, it deviates from the calculated curve due to band repulsion and thereafter both show an opposite dispersion in consistent with calculations. (b) The (111) stop gap does not exhibit any band repulsion due to the absence of second peak in the crossing regime and hence there is no avoided crossing for TM polarization.

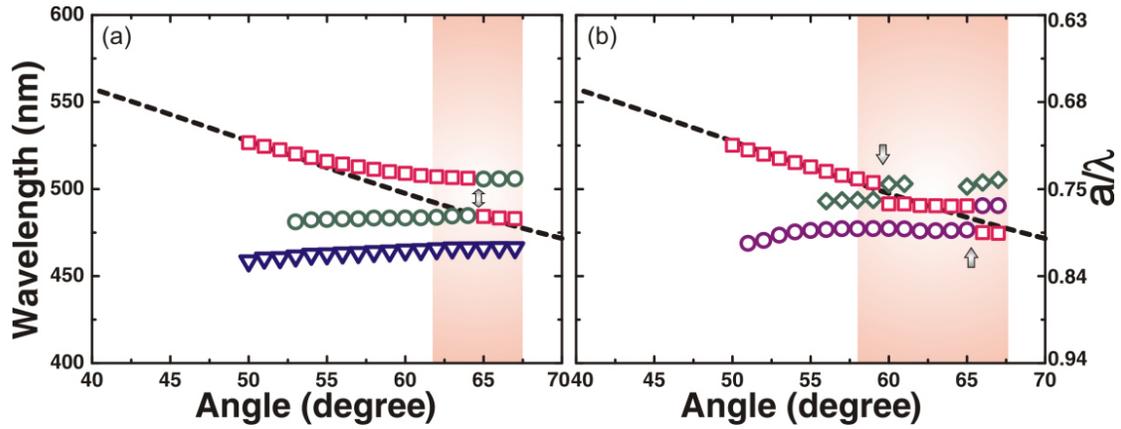

**Fig 6.** The calculated (111) [dashed line] and measured [symbols] stop gap wavelengths along the *W* point for (a) TE and (b) TM polarized light. (a) The measured (111) stop gaps (squares) are in good agreement with calculated (111) stop gap wavelengths. The $S_2$ peak (circles) crosses the (111) stop gap at 65°. The $S_1$ peak (triangles) does not intercept the (111) stop gap in our measurement range. (b) The (111) stop gaps (squares) are in good agreement with its calculated curves. At 59°, the (111) stop gap intercepts the $S_4$ peak (triangles) and exhibit an avoided crossing. The (111) stop gap further encounter the $S_3$ peak (circles) at 65° therein it again shows an avoided crossing.

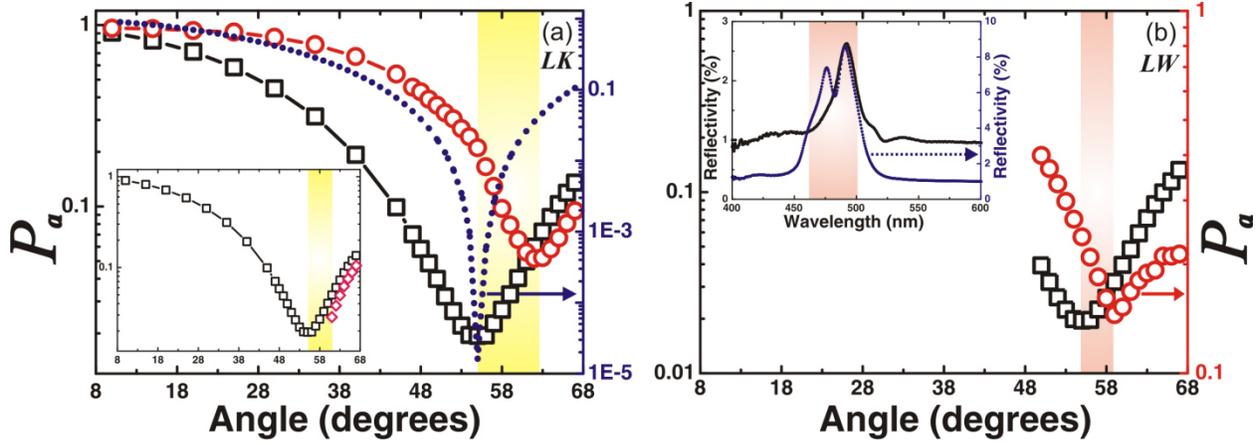

**Fig 7.** Polarization anisotropy ($P_a$) factor as a function of $\theta$ for wave vector shifting towards (a) *K* point and (b) *W* point. The $P_a$ value is shown for long-wavelength limit of 700 *nm* (squares), at the (111) stop gap wavelength (circles), and calculated (dotted lines) value for a film with $n_{eff}$ = 1.436. The minimum value of $P_a$ corresponds to $\theta_B$. The inset in (a) shows the presence of ($\bar{1}11$) stop gap for $\theta > \theta_B$. The inset in (b) indicates the reflectivity spectra at $\theta$ = 62° for wave vector shifting towards the *K* (solid line) and *W* (dotted line) points.